\documentclass{article}

     \usepackage[final]{neurips_2018}

\usepackage[utf8]{inputenc} 
\usepackage[T1]{fontenc}    
\usepackage{hyperref}       
\usepackage{url}            
\usepackage{booktabs}       
\usepackage{amsfonts}       
\usepackage{nicefrac}       
\usepackage{microtype}      
\usepackage{subcaption} 
\usepackage{graphicx}

\usepackage{caption}
\usepackage{amssymb}
\usepackage{graphics}
\usepackage{subcaption}
\usepackage{caption}
\usepackage{subcaption}

\title{Model Size Reduction Using Frequency Based Double Hashing for Recommender Systems}

\author{%
  Caojin Zhang$^*$\\
  Twitter, USA\\
  \texttt{caojinz@twitter.com} \\
   \And
   Yicun Liu\thanks{Both authors contributed equally to this research.}\\
   Twitter, USA\\
  \texttt{yicunl@twitter.com} \\
   \AND
   Yuanpu Xie \\
  Twitter, USA \\
  \texttt{yxie@twitter.com} \\
   \And
  Sofia Ira Ktena \\
  Twitter Inc, UK \\
  \texttt{siraktena@twitter.com} \\ 
  \And
   Alykhan Tejani \\
  Twitter, UK \\
   \texttt{atejani@twitter.com} \\
   \And
   Akshay Gupta\\
   Twitter, UK\\
  \texttt{akshayg@twitter.com} \\
  \And
  Pranay Kumar Myana \\
  Twitter, UK \\
  \texttt{pmyana@twitter.com} \\
  \And 
  Deepak Dilipkumar \\
  Twitter, USA \\
  \texttt{ddilipkumar@twitter.com} \\
  \And
  Suvadip Paul \\
  Twitter, USA \\
  \texttt{suvadipp@twitter.com} \\
  \And
  Ikuhiro Ihara \\
  Twitter, USA \\
  \texttt{iihara@twitter.com} \\
  \And
  Prasang Upadhyaya \\
  Twitter, USA \\
  \texttt{pupadhyaya@twitter.com} \\
  \AND
  Ferenc Huszar \\
  Twitter, UK \\
  \texttt{fhuszar@twitter.com} \\
  \And
  Wenzhe Shi \\
  Twitter, UK \\
  \texttt{wshi@twitter.com} \\
}

\begin{document}
\maketitle

\begin{abstract}
  Deep Neural Networks (DNNs) with sparse input features have been widely used in recommender systems in industry. These models have large memory requirements and need a huge amount of training data. The large model size usually entails a cost, in the range of millions of dollars, for storage and communication with the inference services. In this paper, we propose a hybrid hashing method to combine frequency hashing and double hashing techniques for model size reduction, without compromising performance. We evaluate the proposed models on two product surfaces. In both cases, experiment results demonstrated that we can reduce the model size by around 90 $\%$ while keeping the performance on par with the original baselines.
\end{abstract}

\section{Introduction}

Recommender systems play an important role in a variety of tasks such as display advertising and ranking news feeds for internet companies, having either pointwise click through rate (CTR) prediction or listwise ranking as an objective. Deep learning-based recommendation systems have been widely adopted in industry during the recent years reaching promising performance in different tasks. For instance, Google's wide and deep model \citep{cheng2016wide} led to 3.9 $\%$ improvement of CTR compared to a logistic regression based model in an online setting. According to \citep{Mics}, up to 35 $\%$ of Amazon’s revenue is driven by recommender systems.

\begin{figure}[t!]
    \centering
    \includegraphics[width=0.75\linewidth]{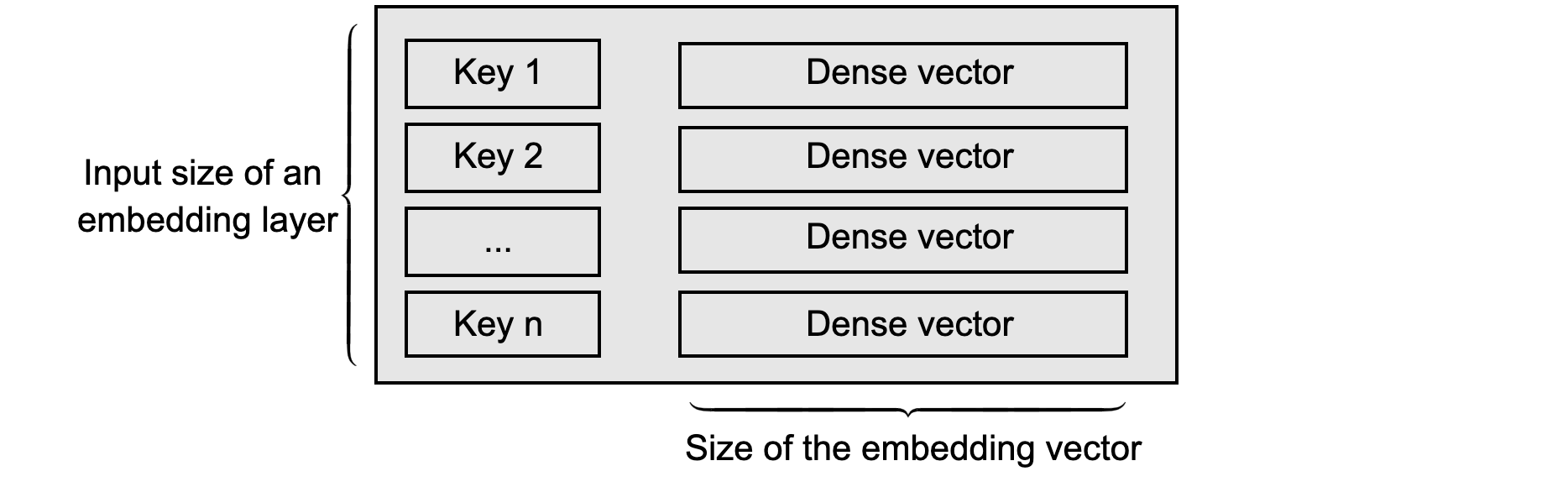}
    \caption{Illustration of an embedding layer, which operates as a lookup table with keys and their corresponding dense vectors}
    \label{fig:incr_embed}
\end{figure}

Unlike many traditional DNN tasks, in deep learning-based recommender systems, user-item features along with their interations are commonly modeled as sparse one-hot or multi-hot vectors. This is due to the prevalence of ID and categorical features. Using a hashing function, each sparse vector is mapped to a key or sparse ID in a look-up table and then transformed into a real-valued dense vector using an embedding layer. In order to minimize collisions, the input size of the embedding layer could be very large, as shown in Fig.~\ref{fig:incr_embed}. As a result, deep learning-based recommendation systems usually contain a large amount of parameters and incur huge memory and communication (I/O) costs for training and inference. For instance, Google's wide and deep model \citep{cheng2016wide} has millions of parameters and requires over 500 billion samples for model training. In \citep{gupta2020architectural}, embedding tables for a single model consume up to tens of GB memory. ByteDance's deep learning-based recommendation system \citep{zhao2020autoemb} adopted a similar architecture as \citep{cheng2016wide} and introduced an extra embedding transformation layer for automatic selection of the embedding vector size. 

In this paper, we propose a hybrid hashing technique for sparse features to minimize the collision rate, while dramatically decreasing the input size of the embedding layers. This means that the model performance can be preserved with significant model size reduction. In section \ref{sec: Related Works}, we review previous works in deep learning-based recommender systems and related hashing techniques. Next, in section \ref{sec: methods} we provide technical details about frequency hashing, double hashing and a seamless way to consolidate them. In section \ref{sec:exp}, we present our experiments on two product surfaces. Finally, we conclude our paper in section \ref{sec:conclusion}. 

The main contribution of the paper includes:
\begin{itemize}
    \item Embedding layers are memory intensive in deep learning-based recommender systems. By reducing the size of embedding layers, we can reduce the large memory capacity challenges in production-scale models. We, thus, propose a hybrid layer to combine frequency hashing with double hashing for sparse input feature embedding, which addresses the pain point of model size in deep learning-based recommender systems.
    
    \item Not all features are of equal importance. Instead of applying double hash functions universally on all features, we decide to eliminate hash collision on the subset of most important features by introducing frequency hashing. 
    
    \item Calculating the hash codes twice using double hashing is computationally expensive, especially in the case of millions of sparse features. With the help of frequency hashing, we only need to apply double hashing on a small amount of low frequency features, which makes the computational complexity comparable to the baseline. 
\end{itemize}

\section{Related Works}\label{sec: Related Works}
\subsection{Sparse Features in Recommendation Systems}

In a production-scale recommendation system, there are usually millions of items and users. For each user, only a small number of items get clicked or viewed. The sparsity and dimensionality of user-item representations are typically high in recommender systems. For instance, \citep{gupta2020architectural} represents user-item interactions as sparse multi-hot vectors. The wide and deep architecture \citep{cheng2016wide} represents user-query history as a sparse high-rank matrix. The deep interest network \citep{zhou2018deep} transforms multi-group categories into sparse binary features. 

Embedding operations are required to transform those sparse feature representations into dense vectors.
However, embedding layers impose challenges in large-scale real-time systems due to their extremely high memory demands. According to \citep{gupta2020architectural}, the size of a production-scale embedding layer could vary between tens of MB to several GB. Moreover, a production-scale recommender system contains 4 to 40 embedding layers. In total, embedding layers in a single production-scale recommender system can require tens of GB memory or more.      
    
\subsection{Hashing Techniques}
A hash function aims at creating mappings from high-dimensional data to low-dimensional representations, where the latter representation is called a hash code. It is widely used for sparse representations such as recommender systems and natural language processing (NLP). A hash function can be either data-independent or learned from data.

Previously, \citep{zhang2018efficient} explored the potential of locality-sensitive hashing (LSH) in recommendation systems. LSH has a very nice property that ensures the probability of collision is much higher for objects that are close to each other than for those that are far away from each other. In other words, this method achieves dimensionality reduction while preserving the proximity of data points in the latent space.
Unlike LSH, which is data-independent, learning to hash provides another alternative that is data-dependent. An overview of learning to hash methods is provided in~\citep{wang2017survey}. \citep{liu2018learning} proposed to learn a hash function in a supervised learning manner for efficient fashion recommendation. Graph network based hashing was recently proposed in \citep{tan2020learning}. \citep{svenstrup2017hash} proposed a hash embedding technique for efficient word representations for NLP. Instead of one hash function, they proposed to apply multiple hash functions to map sparse features to binary representations in a manner that minimizes collisions while decreasing the size of embedding tables. 
Serra et al. \citep{serra2017getting} proposed similar ideas in recommender systems.

\begin{figure*}[t!]
    \centering
    \includegraphics[width=1\linewidth]{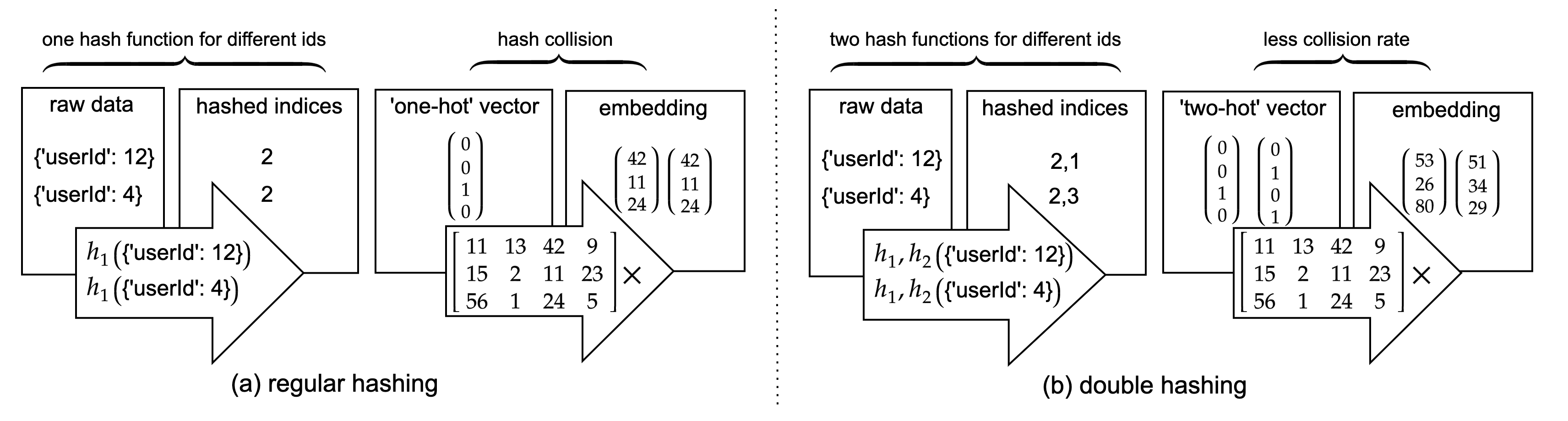}
    \vspace{-0.1in}
    \caption{Illustration of regular hashing method and double hashing method.}
    \label{fig:double_hash}
\end{figure*}

\section{Methods}\label{sec: methods}
\begin{figure*}[t!]
    \centering
    \includegraphics[width=1\linewidth]{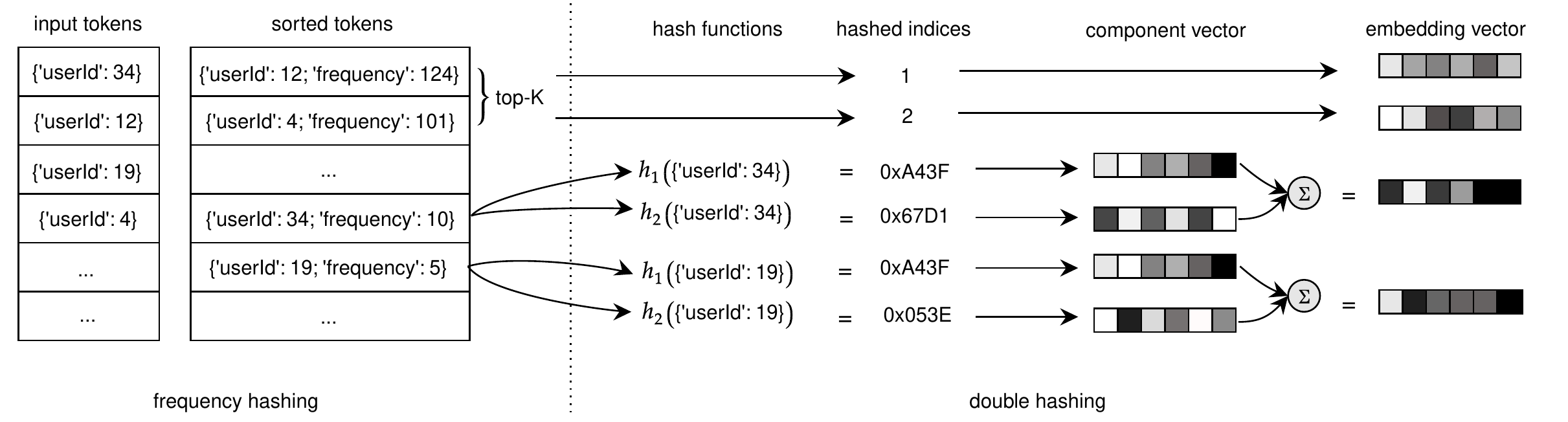}
    \vspace{0.1in}
    \caption{Overview of our proposed hybrid hashing method.}
    \label{fig:hybrid_hash}
\end{figure*}

We propose a hybrid hash embedding layer that adopts techniques from both data-dependent hashing and learning-based hashing, aiming to achieve a trade-off between model size, computational complexity and collision probability. The overall structure of our hybrid layer is illustrated in Fig.~\ref{fig:hybrid_hash}. Our method includes two main components: (1) double hashing (2) frequency hashing.
In this section we first introduce traditional hashing methods as well as some key relevant concepts. Next, we propose to apply double hashing to reduce memory requirements. Finally, we propose frequency hashing to further reduce computational complexity and consolidate it with double hashing. Theoretical analysis further demonstrates that the collision rate decreases with our hybrid method.

\subsection{Hashing Theory}

To create a regular hash embedding for recommender systems, sparse features are randomly mapped to binary codes using a fixed hash function \citep{dang2008recommendation}. Each of these hash codes is mapped to a dense embedding vector, as shown in Fig.~\ref{fig:incr_embed}. We denote the hash function as $h: \mathcal{T} \to \mathcal{B}$, where $\mathcal{T}$ denotes the feature space,  $\mathcal{B}=\{1,\cdots, B\}$ denotes the hashing space and $B$ denotes the total number of permutations in the hash code. To ease our later discussion on model size, we further define $B=2^b$, where $b$ represents the number of bits constructing the hash code. The probability of one or more hash collisions is given by \citep{girault1988generalized} and \citep{suzuki2008birthday}:

\[p_{col}= 1 - \frac{B!}{(B-|\mathcal{T}|)! B^{|\mathcal{T}|}} \approx 1-e^{-|\mathcal{T}|(|\mathcal{T}|-1)/2B} \approx 1-e^{-|\mathcal{T}|^2/(2B)} \]
The expected number of collisions is given by \citep{dt} and \citep{matt}:

\[\mathbb{E}(\textrm{collisions})=|\mathcal{T}|-\mathbb{E}(\textrm{occupied slots})=|\mathcal{T}|-B+\mathbb{E}(\textrm{empty slots})=|\mathcal{T}|-B+B(1-1/B)^{|\mathcal{T}|} \]

where $|\mathcal{T}|$ denotes the total number of features to be hashed. The expected collision rate is given by:
\[r_{col}=\mathbb{E}(\textrm{collision})/B=|\mathcal{T}|/B-1+(1-1/B)^{|\mathcal{T}|}\]

The space complexity of the hash embedding is $O(Bd)=O(B)$, where $d$ is a constant value representing the length of the embedding vector. In terms of time complexity, we consider one hashing operation to cost $O(1)$, so the overall complexity to create the hash embedding is $O(|\mathcal{T}|\ln B)$. In practice, we want $\mathcal{B}$ to be large enough to avoid hash collisions, but the hash embedding size grows exponentially as we increase the number of bits in the hash code.

\subsection{Double Hashing}
To improve the trade-off between collision rate and space complexity, we propose to use double hashing motivated by \citep{svenstrup2017hash}. We apply two independent hash functions $h_1, h_2: \mathcal{T} \to \{1,\cdots,B\}$ to map each sparse feature $f$ to two hash codes $h_1(f), h_2(f)$. Since each hash function can be treated as independent as result of randomization, their combination can be viewed as an approximation of a much larger hash space defined by $h: \mathcal{T} \to  \{1,\cdots,B^2\}$. Instead of the $B^2$ space and computational cost, using this way of hashing only leads to linear cost of $2B$. These hash codes are, then, transformed to a dense feature vector $g(E(h_1(f)), E(h_2(f)))$, where $E$ represents the embedding operation, and $g$ denotes certain aggregation operations include: element-wise summation, or concatenation. 

As illustrated in Fig.~\ref{fig:double_hash}, using two hash functions makes the probability of hash collisions almost negligible even with a much smaller hashing space. This ensures that each feature value has a unique embedding, at the cost of some parameter sharing between feature embeddings. More importantly, this causes less chance of information loss and increases representation power compared to using just one hashing function. 

The main drawback of double hashing is that we have to apply two different hash functions for each sparse feature, which makes model training and inference slower compared to a single hash function as shown in Fig.~\ref{fig:time_comp}. To resolve this issue we introduce frequency hashing.

\subsection{Frequency Hashing}

Empirically, we know some features play more important roles when making a recommendation. Collisions on such features will have unwanted impact. One solution to guarantee the lack of feature collisions is to 
map sparse features to integer IDs $\{1,\cdots, k\}$ using a dictionary. Here, we propose use frequency as a proxy of `feature importance'. We sort sparse features by frequencies and map the top-K frequent sparse features into an ordered set of unique integers IDs $0, \cdots, k-1$. An example of frequency hashing is shown in Fig.~\ref{fig:hybrid_hash}. Since low frequency sparse features are less likely to collide, we can allocate some additional but less hashing space for them. In this way, we can reduce the collision rate (no collisions for most frequent features), while simultaneously decreasing model size and preserving model performance.

The key idea of the hybrid layer described above is to find the top $k$ most frequent features using historical data. Then we apply a frequency hash function (a dictionary), which maps the most frequent features to their unique embedding and transforms the remaining features using the double hashing embedding technique (see Fig.~\ref{fig:hybrid_hash}). With the help of frequency hashing, we only need to apply double hashing on a small amount of low frequency features, which makes the computational complexity comparable to using a single hashing function. This is illustrated in Fig.~\ref{fig:time_comp}.

\subsection{Theoretical Analysis}\label{sec:theory}
We formulate the hash table size, computational complexity and collision rate of the different hashing methods in the table below. This allows a theoretical comparison of the proposed hybrid method with the baseline methods.

\begin{table}[ht]
\scalebox{0.65}{%
\begin{tabular}{c|cccc} \toprule
                 & \textbf{Regular Hashing} & \textbf{Double Hashing} & \textbf{Frequency Hashing} & \textbf{Hybrid Hashing} \\ \midrule
\textbf{Space Complexity} &      $O(B)$        &        $O(B)$        &        $O(|\mathcal{T}|)$          &        $O(B+k)$        \\ 
\textbf{Time Complexity}   & $O(|\mathcal{T}|\ln B)$          &      $O(|\mathcal{T}|\ln B)$          &     $O(|\mathcal{T}|k)$              &            $O((|\mathcal{T}|-k)\ln B +|\mathcal{T}|k)$    \\ 
\textbf{Collision Rate}   &    $|\mathcal{T}|/B-1+(1-1/B)^{|\mathcal{T}|}$             &         $|\mathcal{T}|/B^2-1+(1-1/B^2)^{|\mathcal{T}|}$       &     0              &        $(|\mathcal{T}|-k)/B^2-1+(1-1/B^2)^{|\mathcal{T}|-k}$  \\ \bottomrule   
\end{tabular}}
\label{tab:theoretical_analysis}
\end{table}

To make the comparison more perceivable, suppose our baseline model have feature space $|\mathcal{T}| = 4 M$, $B=2^{22}$. We calculate the expected collision rates when we apply our hybrid hashing method. We report the full collision rate (two hash keys of double hashing are the same) and half collision rate (one of the hash key of double hashing is the same) when we use two independent hash functions and compare them with the case when we only use one hash function (we only have full collision rate).

\begin{table}[ht]
\scalebox{0.67}{%
\begin{tabular}{c|cccc}
    \toprule
    \textbf{Model} &\textbf{Baseline ($100\%$ Size)}& \textbf{Hybrid Hashing ($100\%$ Size)}&\textbf{Hybrid Hashing ($25\%$ Size)} &\textbf{Hybrid Hashing ($10\%$ Size)}\\
    \midrule
    \textbf{Full Collision Rate} & 0.34 & $2.27 \times 10^{-14}$ & $6.27 \times 10^{-12}$ & $1.00 \times 10^{-10}$\\
    \textbf{Half Collision Rate} & N/A & 0.32 & 2.74 & 6.43\\ \bottomrule
    \end{tabular}}
    \label{tab:col_rate}
\end{table}

\section{Experiments}

\subsection{Experiments Setup}
\textbf{Evaluation Metrics} We perform offline experiments to compare the prediction accuracy, computational cost, and model size of the different hashing methods. For prediction accuracy, we choose relative cross entropy (RCE)\citep{ktena2019addressing}. To measure computational cost, we compare the training and inference speed in terms of global steps/second of the Hogwild~\citep{recht2011hogwild} workers using a smoothing factor of $0.995$. With model size we refer to the size of the exported Tensorflow model. 

\textbf{Data \& Models}  
We evaluate our method on two product surfaces, ads prediction and timeline ranking. We use the best production-scale model as a baseline and compare it with our proposed method. For ads prediction, the baseline model architecture is based on the wide and deep model~\citep{cheng2016wide}. For timeline ranking, the baseline model is a multilayer perceptron as a binary classifier to predict user engagement in timeline. Ranking is then performed based on the binary score of the model. 
We simply train the model on the previous day of data and evaluate the model on next day of data. For ads prediction, the training data comprises of 2 days' data, while the following 4 hours are used for testing. To simulate real production services in ads prediction, we also train models on a continuous data stream. In this scenario, we continuously update the model hourly and evaluate it on the data of the next hour. The continuous training and evaluation covers in total a 2-day time period.  

\textbf{Feature Preprocessing} Before feeding into the models, sparse features (categorical and id-related features) are mapped to hash codes using a fixed hash function and then transformed to dense vectors using embedding layers. As for the continuous features, we compute their histograms using historical data and discretize them using bins boundaries that correspond to historical percentiles. Those discrete features are processed similarly to sparse features. 

\textbf{Hyperparamaters} The hyperparameters used for the offline experiments are: stochastic gradient descent optimizer, learning rate 0.001, batch size 128. To find the best hyperparamaters in the hashing configuration, we first identified the best $k$ value for frequency hashing, while still using regular hashing for the rest of features. 
We, subsequently, fixed the frequency hashing part to identify the best number of bins $B$ for double hashing. Each experiment was repeated 5 times and we report here the mean and standard deviation across the different runs.

\begin{table}[t]
\centering
\scalebox{0.75}{%
\begin{tabular}{ccc|ccc}
\toprule
\multicolumn{6}{c}{\textbf{Ads Prediction}}   \\ \midrule
\multicolumn{3}{c|}{\textbf{Frequency Hashing}}        & \multicolumn{3}{c}{\textbf{Frequency Hashing+Double Hashing}}         \\ \midrule
Top K Values & RCE               & Model Size & Input Size Values             & RCE              & Model Size  \\ \midrule
80K          & 11.680$\pm$0.0023 & 29.5\%     & Double Hash $2^{-5}$ & 11.896$\pm$0.010 & 6.6\%      \\
90K          & 11.810$\pm$0.015  & 29.8\%     & Double Hash $2^{-4}$ & 11.949$\pm$0.020 & 8.2\%      \\
100K         & 11.898$\pm$0.021  & 30.1\%     & Double Hash $2^{-3}$ & 11.996$\pm$0.011 & 11.3\%     \\
110K         & 11.941$\pm$0.020  & 30.3\%     & Double Hash $2^{-2}$ & 12.010$\pm$0.012 & 17.8\%     \\
120K         & 11.985$\pm$0.011  & 31.5\%     & One Hash $2^{-1}$    & 11.985$\pm$0.011 & 31.5\%     \\
Baseline     & 11.988$\pm$0.012 & 100\%      & Baseline             & 11.988$\pm$0.011 & 100\%      \\ \bottomrule
\end{tabular}
}
\vspace{0.15in}
\caption{Experiments on configurations of frequency hashing and double hashing on ads prediction.}
\label{tab:results-ads}
\end{table}

\begin{table}[t]
\centering
\scalebox{0.75}{%
\begin{tabular}{ccc|ccc}
\toprule
\multicolumn{6}{c}{\textbf{Timeline Ranking}}              \\ \midrule
\multicolumn{3}{c|}{\textbf{Frequency Hashing}}  & \multicolumn{3}{c}{\textbf{Frequency Hashing+Double Hashing}}              \\ \midrule
Top K Values             & RCE   & Model Size & Input Size Values             & RCE   & Model Size \\ \midrule
80K & 17.441$\pm$0.073 & 22.7\%   & Double Hash $2^{-5}$ & 17.521$\pm$0.074 & 12.7\%      \\
90K & 17.486$\pm$0.013 & 22.8\%    & Double Hash $2^{-4}$ & 17.524$\pm$0.059 & 13.3\%      \\
100K & 17.511$\pm$0.089 & 22.9\%    & Double Hash $2^{-3}$ &17.510$\pm$0.041 & 14.7\%      \\
110K & 17.471$\pm$0.081 & 23.0\%    & Double Hash $2^{-2}$ & 17.523$\pm$0.035 & 17.4\%      \\
120K & 17.473$\pm$0.032 & 23.1\%    & One Hash $2^{-1}$ & 17.511$\pm$0.089 & 22.9\%      \\
Baseline & 17.529$\pm$0.050 & 100\%   & Baseline & 17.529$\pm$0.050 & 100\%    \\ \bottomrule
\end{tabular}
}
\vspace{0.15in}
\caption{Experiments on configurations of frequency hashing and double hashing on timeline ranking.}
\label{tab:results-tq}
\end{table}

\subsection{Experiments Results}\label{sec:exp}
We perform our experiments on ads prediction model and timeline ranking model. As described above, we experiment with different sizes of the frequency hashing layer while keeping the input size fixed (half of the original input size) for the rest of features. We show comparative results in Table~\ref{tab:results-ads} and Table~\ref{tab:results-tq}. We report the evaluation RCE while the input and model sizes are normalized with respect to the baseline. 

\begin{figure}[h]
  \centering
  \begin{minipage}[c]{.47\textwidth}
  \centering
  \includegraphics[width=1\textwidth]{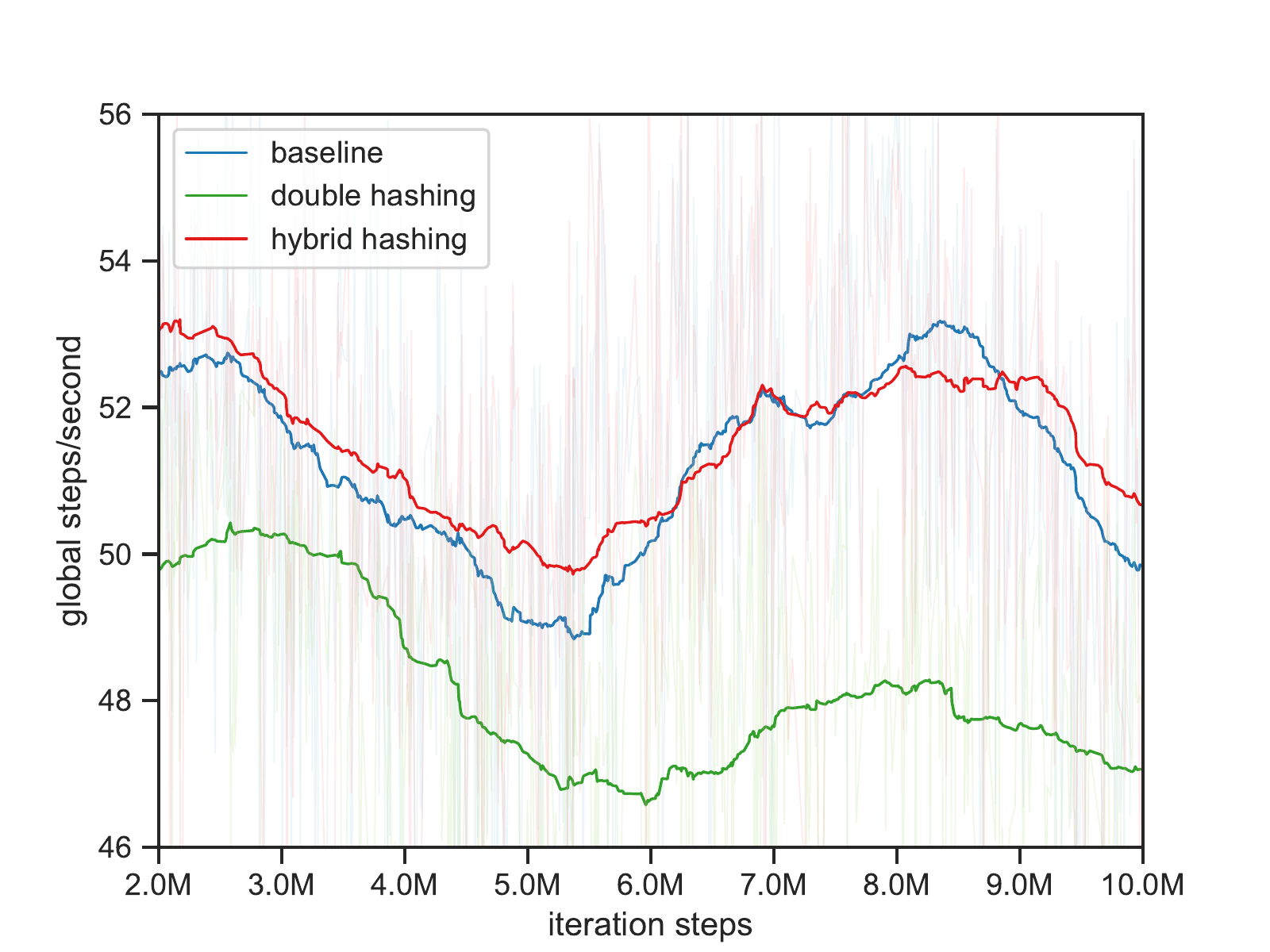}
  \vspace{-0.1in}
  \caption{Comparison of training time.  
  For our experiments, double hashing is roughly $22\%$ slower than the baseline. Our hybrid method is at least comparable to, or slightly better than the baseline.}
  \label{fig:time_comp}
  \end{minipage}
  \hspace{0.2in}
  \begin{minipage}[c]{.47\textwidth}
  \centering
  \includegraphics[width=1\textwidth]{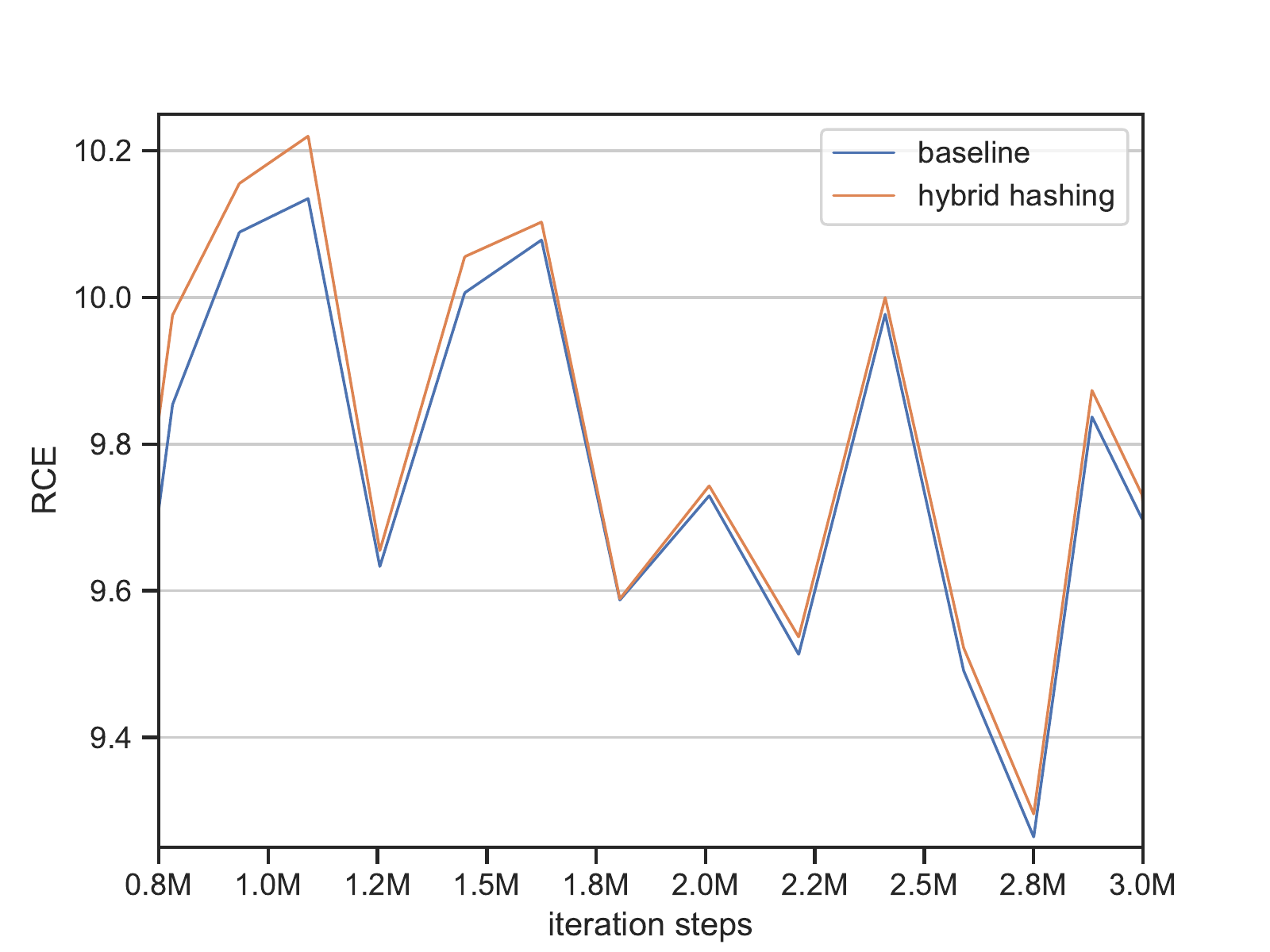}
  \vspace{-0.1in}
  \caption{Comparison of RCE during continuous training. The hybrid hashing method performs consistently
better than (or at worst on par with) the baseline in terms of prediction accuracy.}
  \label{fig:online_comp}
  \end{minipage}
\end{figure}

For ads prediction, frequency hashing with $k=120K$ can reduce the model size to $31\%$ while preserving the model performance on par with the baseline. By further increasing the value of $k$, the gains reach a plateau. Then we fix $k=120K$ and explore what the optimal input size $b$ for the double hashing layer is in table \ref{tab:results-ads}. Since double hashing can decrease collision rates with a much smaller hashing size, we can further reduce the model size to $11.3\%$ of the baseline model, without compromising model performance. Additionally, we use the best model from the previous experiments and train it on a continuous data stream. We compare it with the baseline (production) model in Fig.~\ref{fig:online_comp}. Results indicate that the proposed method can achieve a marginal but consistent RCE gain compared to the baseline model.

For timeline ranking, we observe that frequency hashing with $k=100K$ can reduce the model size to $23\%$ with RCE very close to baseline in table \ref{tab:results-tq}. When we further increase the value of $k$, we observe that performance reaches a plateau. This is possibly due to the majority of model sparsity comes from id-related features, where over-memorization of the ids could introduce potential overfitting. Then we fix $k=100K$ and explore different number of bits $b$ for double hashing. We further reduce the model size to $12.7\%$, while maintaining the model performance comparable to baseline model.

The main drawback of using double hashing is the increase in computational time. Two different hash functions need to be called for each feature, which slows down both model training and inference. To get a better estimate of time performance, we compare the number of global steps per second for the baseline model, the model that uses pure double hashing, and the model that uses frequency hashing and double hashing under the same computational constraints in Fig.~\ref{fig:time_comp}. This illustrates that the most expensive method is double hashing followed by the baseline and hybrid hashing. We observe an almost $22\%$ decrease in the number of global steps per second for the double hashing method compared to the baseline. When frequency hashing is combined with double hashing, the computational time becomes comparable to the baseline.

\section{Conclusion}\label{sec:conclusion}
In this paper, we propose to combine frequency hashing with double hashing for model size reduction. Not only does it dramatically reduce the memory requirements in deep learning-based recommender systems, but also decreases the computational cost compared to double hashing alone. Our frequency-based hash function is learned offline using historical data. In future research, we aim to study how to perform dynamic updates of the frequency hashing layer to better support warm starting or continuous learning scenarios. In addition, we can explore feature importance measures other than frequency, e.g. entropy, Fisher Information \citep{theis2018faster}, or Gini importance \citep{menze2009comparison} to increase the performance of the compressed model.

\bibliographystyle{plainnat}
\bibliography{nips_2018}

\begin{thebibliography}{20}
\providecommand{\natexlab}[1]{#1}
\providecommand{\url}[1]{\texttt{#1}}
\expandafter\ifx\csname urlstyle\endcsname\relax
  \providecommand{\doi}[1]{doi: #1}\else
  \providecommand{\doi}{doi: \begingroup \urlstyle{rm}\Url}\fi

\bibitem[Cheng et~al.(2016)Cheng, Koc, Harmsen, Shaked, Chandra, Aradhye,
  Anderson, Corrado, Chai, Ispir, et~al.]{cheng2016wide}
Heng-Tze Cheng, Levent Koc, Jeremiah Harmsen, Tal Shaked, Tushar Chandra,
  Hrishi Aradhye, Glen Anderson, Greg Corrado, Wei Chai, Mustafa Ispir, et~al.
\newblock Wide and deep learning for recommender systems.
\newblock In \emph{Proceedings of the 1st workshop on deep learning for
  recommender systems}, pages 7--10, 2016.

\bibitem[College(2018)]{dt}
Dartmouth College.
\newblock Probability calculations in hashing, 2018.
\newblock URL
  \url{https://math.dartmouth.edu/archive/m19w03/public_html/Section6-5.pdf}.

\bibitem[Dang(2008)]{dang2008recommendation}
Quynh Dang.
\newblock \emph{Recommendation for applications using approved hash
  algorithms}.
\newblock US Department of Commerce, National Institute of Standards and
  Technology, 2008.

\bibitem[Girault et~al.(1988)Girault, Cohen, and
  Campana]{girault1988generalized}
Marc Girault, Robert Cohen, and Mireille Campana.
\newblock A generalized birthday attack.
\newblock In \emph{Workshop on the Theory and Application of of Cryptographic
  Techniques}, pages 129--156. Springer, 1988.

\bibitem[Gupta et~al.(2020)Gupta, Wu, Wang, Naumov, Reagen, Brooks, Cottel,
  Hazelwood, Hempstead, Jia, et~al.]{gupta2020architectural}
Udit Gupta, Carole-Jean Wu, Xiaodong Wang, Maxim Naumov, Brandon Reagen, David
  Brooks, Bradford Cottel, Kim Hazelwood, Mark Hempstead, Bill Jia, et~al.
\newblock The architectural implications of facebook's dnn-based personalized
  recommendation.
\newblock In \emph{2020 IEEE International Symposium on High Performance
  Computer Architecture (HPCA)}, pages 488--501. IEEE, 2020.

\bibitem[Ktena et~al.(2019)Ktena, Tejani, Theis, Myana, Dilipkumar, Husz{\'a}r,
  Yoo, and Shi]{ktena2019addressing}
Sofia~Ira Ktena, Alykhan Tejani, Lucas Theis, Pranay~Kumar Myana, Deepak
  Dilipkumar, Ferenc Husz{\'a}r, Steven Yoo, and Wenzhe Shi.
\newblock Addressing delayed feedback for continuous training with neural
  networks in ctr prediction.
\newblock In \emph{Proceedings of the 13th ACM Conference on Recommender
  Systems}, pages 187--195, 2019.

\bibitem[Liu et~al.(2018)Liu, Du, Zhu, Shen, and Huang]{liu2018learning}
Luyao Liu, Xingzhong Du, Lei Zhu, Fumin Shen, and Zi~Huang.
\newblock Learning discrete hashing towards efficient fashion recommendation.
\newblock \emph{Data Science and Engineering}, 3\penalty0 (4):\penalty0
  307--322, 2018.

\bibitem[Menze et~al.(2009)Menze, Kelm, Masuch, Himmelreich, Bachert, Petrich,
  and Hamprecht]{menze2009comparison}
Bjoern~H Menze, B~Michael Kelm, Ralf Masuch, Uwe Himmelreich, Peter Bachert,
  Wolfgang Petrich, and Fred~A Hamprecht.
\newblock A comparison of random forest and its gini importance with standard
  chemometric methods for the feature selection and classification of spectral
  data.
\newblock \emph{BMC bioinformatics}, 10\penalty0 (1):\penalty0 213, 2009.

\bibitem[Might(2018)]{matt}
Matt Might.
\newblock Counting hash collisions with the birthday paradox, 2018.
\newblock URL \url{http://matt.might.net/articles/counting-hash-collisions/}.

\bibitem[Recht et~al.(2011)Recht, Re, Wright, and Niu]{recht2011hogwild}
Benjamin Recht, Christopher Re, Stephen Wright, and Feng Niu.
\newblock Hogwild: A lock-free approach to parallelizing stochastic gradient
  descent.
\newblock In \emph{Advances in neural information processing systems}, pages
  693--701, 2011.

\bibitem[Serr{\`a} and Karatzoglou(2017)]{serra2017getting}
Joan Serr{\`a} and Alexandros Karatzoglou.
\newblock Getting deep recommenders fit: Bloom embeddings for sparse binary
  input/output networks.
\newblock In \emph{Proceedings of the Eleventh ACM Conference on Recommender
  Systems}, pages 279--287, 2017.

\bibitem[Suzuki et~al.(2008)Suzuki, Tonien, Kurosawa, and
  Toyota]{suzuki2008birthday}
Kazuhiro Suzuki, Dongvu Tonien, Kaoru Kurosawa, and Koji Toyota.
\newblock Birthday paradox for multi-collisions.
\newblock \emph{IEICE Transactions on Fundamentals of Electronics,
  Communications and Computer Sciences}, 91\penalty0 (1):\penalty0 39--45,
  2008.

\bibitem[Svenstrup et~al.(2017)Svenstrup, Hansen, and
  Winther]{svenstrup2017hash}
Dan~Tito Svenstrup, Jonas Hansen, and Ole Winther.
\newblock Hash embeddings for efficient word representations.
\newblock In \emph{Advances in Neural Information Processing Systems}, pages
  4928--4936, 2017.

\bibitem[Tan et~al.(2020)Tan, Liu, Zhao, Yang, Zhou, and Hu]{tan2020learning}
Qiaoyu Tan, Ninghao Liu, Xing Zhao, Hongxia Yang, Jingren Zhou, and Xia Hu.
\newblock Learning to hash with graph neural networks for recommender systems.
\newblock In \emph{Proceedings of The Web Conference 2020}, pages 1988--1998,
  2020.

\bibitem[Theis et~al.(2018)Theis, Korshunova, Tejani, and
  Husz{\'a}r]{theis2018faster}
Lucas Theis, Iryna Korshunova, Alykhan Tejani, and Ferenc Husz{\'a}r.
\newblock Faster gaze prediction with dense networks and fisher pruning.
\newblock \emph{arXiv preprint arXiv:1801.05787}, 2018.

\bibitem[Wang et~al.(2017)Wang, Zhang, Sebe, Shen, et~al.]{wang2017survey}
Jingdong Wang, Ting Zhang, Nicu Sebe, Heng~Tao Shen, et~al.
\newblock A survey on learning to hash.
\newblock \emph{IEEE transactions on pattern analysis and machine
  intelligence}, 40\penalty0 (4):\penalty0 769--790, 2017.

\bibitem[Xie et~al.(2018)Xie, Lian, Liu, Wang, Wu, Wang, and Chen]{Mics}
Xing Xie, Jianxun Lian, Zheng Liu, Xiting Wang, Fangzhao Wu, Hongwei Wang, and
  Zhongxia Chen.
\newblock Personalized recommendation systems: Five hot research topics you
  must know, 2018.
\newblock URL
  \url{https://www.microsoft.com/en-us/research/lab/microsoft-research-
  asia/articles/personalized- recommendation- systems/}.

\bibitem[Zhang et~al.(2018)Zhang, Fan, and Wang]{zhang2018efficient}
Kunpeng Zhang, Shaokun Fan, and Harry~Jiannan Wang.
\newblock An efficient recommender system using locality sensitive hashing.
\newblock In \emph{Proceedings of the 51st Hawaii International Conference on
  System Sciences}, 2018.

\bibitem[Zhao et~al.(2020)Zhao, Wang, Chen, Zheng, Liu, and
  Tang]{zhao2020autoemb}
Xiangyu Zhao, Chong Wang, Ming Chen, Xudong Zheng, Xiaobing Liu, and Jiliang
  Tang.
\newblock Autoemb: Automated embedding dimensionality search in streaming
  recommendations.
\newblock \emph{arXiv preprint arXiv:2002.11252}, 2020.

\bibitem[Zhou et~al.(2018)Zhou, Zhu, Song, Fan, Zhu, Ma, Yan, Jin, Li, and
  Gai]{zhou2018deep}
Guorui Zhou, Xiaoqiang Zhu, Chenru Song, Ying Fan, Han Zhu, Xiao Ma, Yanghui
  Yan, Junqi Jin, Han Li, and Kun Gai.
\newblock Deep interest network for click-through rate prediction.
\newblock In \emph{Proceedings of the 24th ACM SIGKDD International Conference
  on Knowledge Discovery \& Data Mining}, pages 1059--1068, 2018.

\end{thebibliography}

\end{document}